\documentclass[aps,prl,reprint,superscriptaddress]{revtex4-2}
\usepackage{graphicx,amsmath,siunitx,mathtools}
\usepackage[colorlinks=true]{hyperref}

\begin{document}

\title{Exactly Solvable Model of Random Walks with Stochastic Exchange}

\author{Jos\'e Julian D\'{\i}az-P\'erez}
\affiliation{Department of Functional Analysis, Faculty of Mathematics and Computer Science, University of Havana, CP 10400, La Habana, Cuba}
\affiliation{Group of Complex Systems and Statistical Physics, Faculty of Physics, University of Havana, CP 10400, La Habana, Cuba}
\author{Roberto Mulet}
\affiliation{Group of Complex Systems and Statistical Physics and Department of Theoretical Physics, Faculty of Physics, University of Havana, CP 10400, La Habana, Cuba}
\affiliation{The Clinical Hospital of Chengdu Brain Science Institute, MOE Key Lab for Neuroinformation, School of Life Science and Technology, University of Electronic Science and Technology of China, Chengdu, China}
\email{roberto.mulet@gmail.com}
\date{\today}

\begin{abstract}
  We solve exactly the non-equilibrium dynamics of two discrete random walkers moving in channels with transition rates $p \neq q$ that swap positions at a rate $s$. We compute exactly the joint probability distribution $P_{n,m}(t)$ for the walkers, revealing the existence of two dynamical crossovers. The first signals the passage from independent diffusion to a swap-dominated regime where the particles act as identical random walkers swapping positions. The second crossover occurrs when both channels become indistinguishable and the walkers move around the same position. Furthermore, we demonstrate the existence of a persistent spatial anisotropy defined by the difference between the second moments of the probability distributions in the two channels. Our results may provide a quantitative framework to understand diverse systems. In biology, it is motivated by motor proteins (kinesin/dynein) exchanging cargo leadership,  membrane receptors swapping binding partners, or brain synapses with activity-dependent plasticity. In finance, it models traders with distinct risk profiles swapping positions in limit-order books, or volatility spillover between coupled markets. These diverse systems share a unifying theme:  exchange processes mediate macroscopic correlations despite individual heterogeneity.
\end{abstract}

\maketitle

The non-equilibrium dynamics of interacting particles is a central theme in modern statistical physics\cite{Caceres,Krapivsky10}. A fundamental challenge in this domain is understanding how microscopic interactions—especially among heterogeneous agents— generate correlations that shape macroscopic phenomena. In particular, the study of particles constrained to move in one dimension offers a rich testing ground for the principles of statistical mechanics and helps to build basic models for real-world applications. Models of interacting particles\cite{Song97, Kumar08,Nelissen_2007}, exclusion processes\cite{Derrida93,Derrida04}, and one-dimensional systems in the presence of resetting\cite{Evans11,Evans2020,Biroli24} have allowed the derivation of exact or near-exact results, which have helped  to gain theoretical insights about many non-equilibrium phenomena. However, there is an important class of models that remains essentially unexplored: those in which particles swap positions while moving in one-dimensional channels. We find this striking, considering the many real-world situations that can motivate such models.



For example, in biological systems, motor proteins coordinate cargo transport along microtubules through discrete, energy-driven steps\cite{VALE,Melani08}. Experiments have shown that kinesin (fast, $q$) and dynein (slow, $p$) alternate cargo transport roles when bound to a shared cargo, enabling coordinated motion\cite{Welte}. Similarly, during learning and memory formation, synapses exhibit activity-dependent plasticity, where  each synapse undergoes stochastic strengthening or weakening. These processes can be modeled as random walks at distinct rates $p$ and $q$, which represent the individual responsiveness of the synapses to neural activity\cite{Suri,babadi}. Occasionally, these synapses exchange limiting molecular components — such as signaling molecules or receptor pools — at a finite rate $s$, akin to a swap mechanism\cite{Hanus}. A similar dynamics occurs in financial systems\cite{Kyle, Bouchaud2004,Kevin}, in this case the  distinct intrinsic volatilities of two markets correspond to the transition rates $q$ and $p$, while the leverage-induced rebalancing plays the role of the swap rate $s$, mediating the dynamic redistribution of risk. The model also describes traders with different risk profiles $q$ and $p$ swapping positions in limit order books. Capturing this exchange phenomenon requires models that go beyond the standard single-particle descriptions or hard-core interactions. 

\begin{figure}[htbp]
	\centering
	\includegraphics[width=0.47\textwidth]{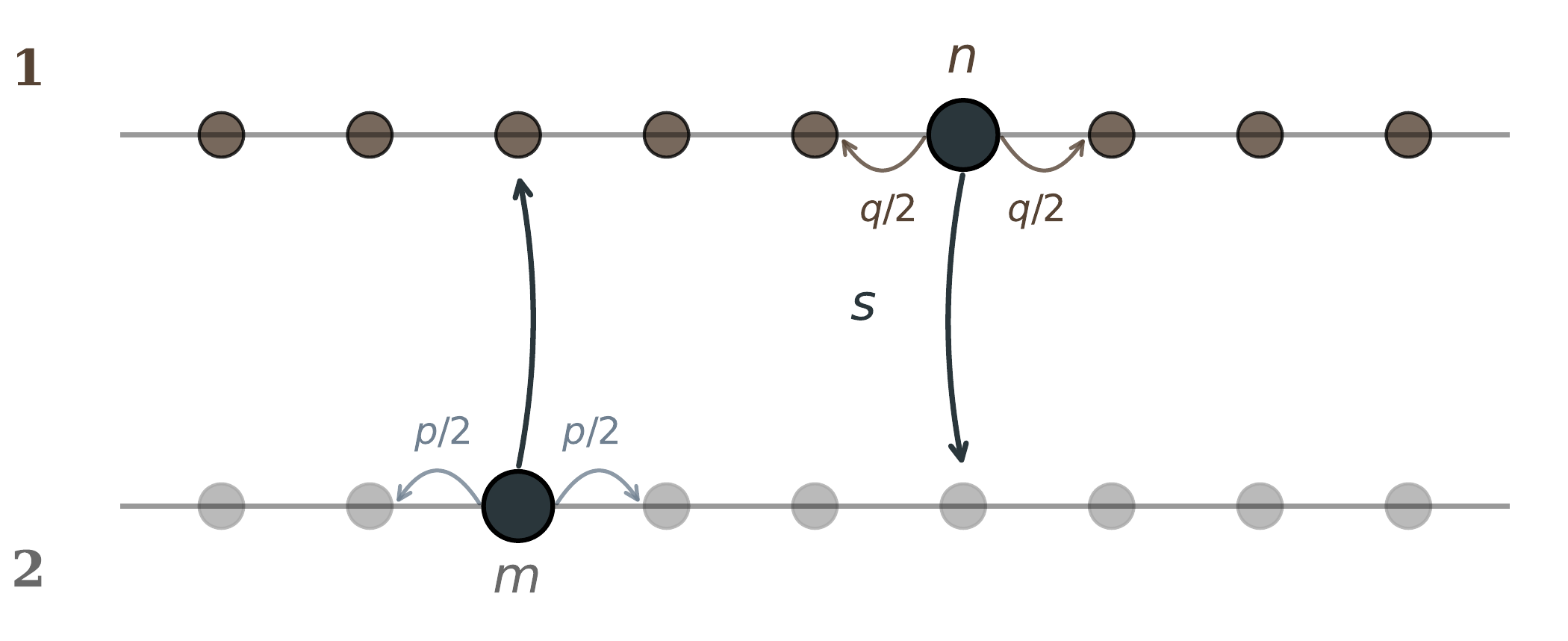}
	\caption{Schematic representation of the model: $q$ and $p$ are the transition rates and $n$ and $m$ represent the positions of the particles in the upper and lower channels, respectively. $s$ is the swapping rate between the particles. }
	\label{fig:diagram}
\end{figure}

In these scenarios, we hypothesize that local exchange processes generate correlations that govern the macroscopic behavior of the system. Motivated by the examples above, we consider two indistinguishable random walkers with transition rates $q$ and $p$ moving on discrete one-dimensional lattices. They exchange positions at rate $s$: if one particle occupies position $n$ in channel one, and the other $m$ in channel two, they swap to $m$ and $n$ at rate $s$ (see Figure \ref{fig:diagram}). Equivalently, this describes a single particle moving in two dimensions with anisotropic transition rates $q$ and $p$ {\em swapping coordinates at rate $s$}.

The master equation governing the dynamics is: 
\begin{align}
  \frac{dP_{n,m}(t)}{dt} &= \frac{q}{2} \sum_{\sigma=\pm1} P_{n+\sigma,m}(t) + \frac{p}{2} \sum_{\sigma=\pm1} P_{n,m+\sigma}(t) \nonumber \\
  &+ s P_{m,n}(t) - (q+p+s) P_{n,m}(t)
  \label{eq:ME2P}
\end{align}
where the first two terms describe the independent hopping of the particles, the third term represents position swapping, and the last ensures normalization. We assume initial positions $P_{n,m}(t=0) = \delta_{n,n_0}\delta_{m,m_0}$.

Applying the Fourier transform $G(k_x,k_y,t)= \sum_{n,m} e^{i k_x n + i k_y m} P_{n,m}(t)$ with $\vec{k} = (k_x,k_y)$ and $\underline{\vec{k}} \equiv (k_y,k_x)$, followed by the Laplace transform $\hat{G}(\vec{k},r) = \int_0^\infty e^{-rt} G(\vec{k},t) dt$, equation (\ref{eq:ME2P}) transforms into:
\begin{equation}
  r \hat{G}(\vec{k},r) - \hat{G}(\vec{k},0) = w(\vec{k}) \hat{G}(\vec{k},r) + s \hat{G}(\underline{\vec{k}},r)
  \label{eq:Equation2D-Laplace}
\end{equation}
where  $w(\vec{k}) = q(\cos k_x - 1) + p(\cos k_y - 1) - s$. Notice that (\ref{eq:Equation2D-Laplace}) is actually a system of equations coupling $\hat{G}(\vec{k},r)$ and $\hat{G}(\underline{\vec{k}},r)$.  We solve this system by the method of successive iterations (see Supplementary Material) and obtain:
\begin{equation}
  \hat{G}(\vec{k},r) = \frac{(r - w(\underline{\vec{k}}))\hat{G}(\vec{k},0) + s\hat{G}(\underline{\vec{k}},0)}{(r - r_1)(r - r_2)}
  \label{eq:sol2D-Laplace-b}
\end{equation}
where $r_{1,2}=\frac{ w(\vec{k})+w(\underline{\vec{k}})}{2} \pm \frac{1}{2} \sqrt{( w(\vec{k}) - w(\underline{\vec{k}}))^2 + 4s^2}$ are the eigenvalues. 

Inverting (\ref{eq:sol2D-Laplace-b}) to the time domain results in:
\begin{equation}
  G(\vec{k},t) = \frac{A e^{r_1 t} - B e^{r_2 t}}{r_1 - r_2}
  \label{eq:G(k,t)}
\end{equation}
where $A = s G(\underline{\vec{k}},0) + (r_1 - w(\underline{\vec{k}}))G(\vec{k},0)$ and $B = s G(\underline{\vec{k}},0) + (r_2 - w(\underline{\vec{k}}))G(\vec{k},0)$.

The derivation of the full probability distribution, involves a non-trivial inversion of  eq.(\ref{eq:G(k,t)}). The inversion combines, proper regularization of integrals, series expansions of special functions and the integral representation of the Dirac delta function. It appears in detail in the Supplementary Material. The final expression (\refeq{eq:fullsolution}), albeit involved, is exact and amenable to asymptotic analysis.
\begin{equation}
\begin{split}
  P_{n,m}(t) &= e^{-(q+p+s)t} \Bigg[ I_{n-n_0}(qt) I_{m-m_0}(pt) \\
  &+ \frac{st}{4} \sum_{\sigma=\pm 1} \Bigg( \int_0^1 \frac{dz}{\sqrt{z}} \mathcal{I}_\sigma(z) I_{n-n_0}(\Theta_\sigma^+) I_{m-m_0}(\Theta_\sigma^-) \\
  &+ \int_0^1 \frac{dz}{\sqrt{z}} I_0(st \sqrt{1-z}) I_{n-m_0}(\Theta_\sigma^+) I_{m-n_0}(\Theta_\sigma^-) \Bigg) \Bigg]
\end{split}
\label{eq:fullsolution}
\end{equation}
where $\mathcal{I}_\sigma(z) = \frac{1 + \sigma \sqrt{z}}{\sqrt{1-z}} I_1(st \sqrt{1-z})$, $\Theta_\sigma^{\pm} = \frac{q+p}{2} t \pm \sigma \frac{t\sqrt{z}}{2} (q-p)$, and $I_\nu(\cdot)$ are modified Bessel functions.

Equation (\ref{eq:fullsolution}) exhibits three distinct dynamical regimes. For short times or slow swapping rates ($t \ll 1/s$), we have the \textit{independent diffusion} regime $ P_{n,m}(t) \approx e^{-(q+p)t} I_{n-n_0}(qt)\, I_{m-m_0}(pt),$ in which each particle diffuses with its intrinsic rate ($q$ or $p$), retaining its mobility. At larger times  ($t \gg 1/s$), the system enters the \textit{swap-dominated} regime, where
\begin{eqnarray}\nonumber
P_{n,m}(t) &\approx& \frac{e^{-(q+p)t}}{2} \Big[ I_{n - n_0}\!\left( \tfrac{q + p}{2}t \right) I_{m - m_0}\!\left( \tfrac{q + p}{2}t \right) \\&+& I_{n - m_0}\!\left( \tfrac{q + p}{2}t \right) I_{m - n_0}\!\left( \tfrac{q + p}{2}t \right) \Big],
\end{eqnarray}
(see Supplementary Material for the derivation). In this regime, the particles in both channels change their positions at a rate $s$ from $\langle n_0 \rangle $ to $\langle m_0 \rangle$ diffusing with an effective rate $(q+p)/2$. The distribution in each channel is bimodal. Finally, in the \textit{mixing} regime, the particles become indistinguishable and move like standard random walkers around a fixed position $c = \frac{n_0 + m_0}{2}$. A simple estimate for the mixing time follows from noting that, for $t \gg 1/s$, the variance of the particles around each initial position is $\frac{q+p}{2}t$. The overlap between the distributions distinguishing these peaks occurs when this matches the initial separation $n_0 - m_0$, giving $ t_{m} \sim \frac{4 (n_0 - m_0)^2}{q+p} + \frac{1}{s}$. A more formal asymptotic analysis can be found in the Supplementary Material. 

\begin{figure}[!htbp]
	\centering
	\includegraphics[width=0.5\textwidth]{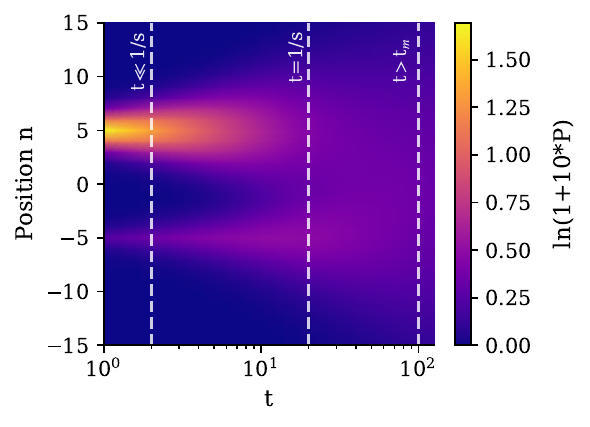}
        \includegraphics[width=0.5\textwidth]{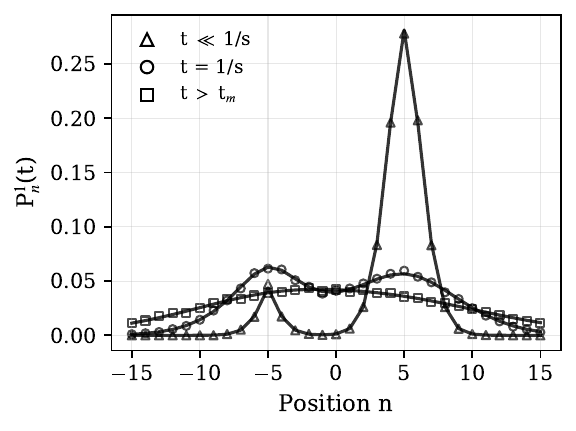}
	\caption{ (Color online) (a) Evolution of the marginal distribution $P_n(t)$ with a particle starting in channel one at $n_0=5$ and a particle starting in channel two at $m_0=-5$. (b) Snapshots at different instants of time of $P_n(t)$. Parameters: $p=0.2, q=2.0, s=0.1$. Dashed lines indicate initial conditions. The logarithmic scale enhances visibility of low-probability regions. The vertical lines indicate the crossover times defined in the main text. }
	\label{fig:dynamics}
\end{figure}

These three regimes are shown in  Fig \ref{fig:dynamics} where we plot the evolution of the marginalized distribution function, $P(n,t)$, computed from \eqref{eq:fullsolution} as a function of time. In the figure it is easily seen that thedistribution starts concentrated at $n_0=5$, splits at already $t \sim 1/s$ and these two peaks spread until they collapse at $t\sim t_m$ at the central point. 

\begin{figure}[!htbp]
	\centering
	\includegraphics[width=0.49\textwidth]{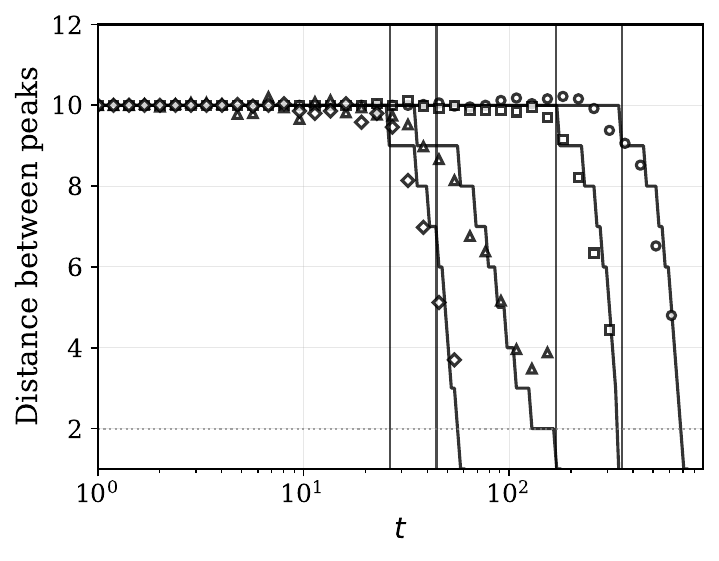}
	\caption{ Distance between the greatest peak in exact and long-time approximation as a function of time. The plot shows the absolute difference in peak positions between the exact marginal distribution $P_n^1(t)$ and its long-time approximation for different parameter sets: $q=0.2, s=0.05$, $q=0.5, s=0.5$ (dark gray), $q=4.0, s=0.05$, and $q=4.0, s=0.5$, with $p=0.1$, $n_0=10$, $m_0=-10$ for all cases. Solid lines represent theoretical predictions, while markers show Monte Carlo simulation results. Vertical lines mark the critical mixing times $t_{m} \sim 4(n_0-m_0)^2/(q+p) + 1/s$ for each parameter set. }
	\label{fig:due_pick}
\end{figure}

Figure \ref{fig:due_pick} compares our estimate of $t_{m}$ with numerical simulations. We show the distance between the maxima of the exact (bimodal) marginal in the swapping regime and the long-time solution in the mixed regime. The curves confirm that the change from a bimodal to a unimodal profile occurs just after $t_m$, in agreement with the theory.
The mean position of a particle in both channels converges rapidly (with a characteristic timescale $t \sim 1/s$ ) to $\frac{n_0 + m_0}{2}$ suggesting that the swapping rate acts as an effective attractive force between the particles.

The position covariance is given by
\begin{equation}
  \langle n(t)m(t) \rangle_c = -\frac{(n_0 - m_0)^2}{4} \left(1 - e^{-4st}\right).
\end{equation}
It evolves exponentially fast, starting from zero, toward its (negative) asymptotics value signaling the existence of a long time anti-correlation between the particle positions proportional to the square of their initial separation. It can be interpreted considering that the particles in the channels tend to occupy positions on the opposite sides of their collective center of mass $\frac{n_0+m_0}{2}$. This swap-induced anti-correlation underlies how this swapping  mechanism may enhance the coordinated transport in motor proteins, and also the risk redistribution in coupled financial markets.

The second moments computed in both channels exhibit a richer dynamics: \begin{eqnarray}
	\langle n^2(t),m^2(t) \rangle_v &=& \frac{q+p}{2}t +\frac{(n_0 - m_0)^2}{4} (1 - e^{-4st})  \nonumber \\
	&\pm &\frac{q-p}{4s} (1 - e^{-2st})
	\label{eq:variances}
\end{eqnarray}
where the $+,(-)$ correspond to the $n^2$ and $m^2$ respectively. At intermediate times around $t \sim 1/(4s)$ and $t \sim 1/(2s)$, two competing mechanism emerge: the contribution from the initial separation between the particles $\frac{(n_0 - m_0)^2}{4}$ and the mobility-dependent term $\frac{q-p}{2s}$. However, as time increases, both exponential terms become constant and negligible compared to the linear collective diffusion term $\frac{q+p}{2}t$, which ultimately dominates the long-time behavior in both channels.
 
\begin{figure}[!htbp]
  \centering
  \includegraphics[width=0.49\textwidth]{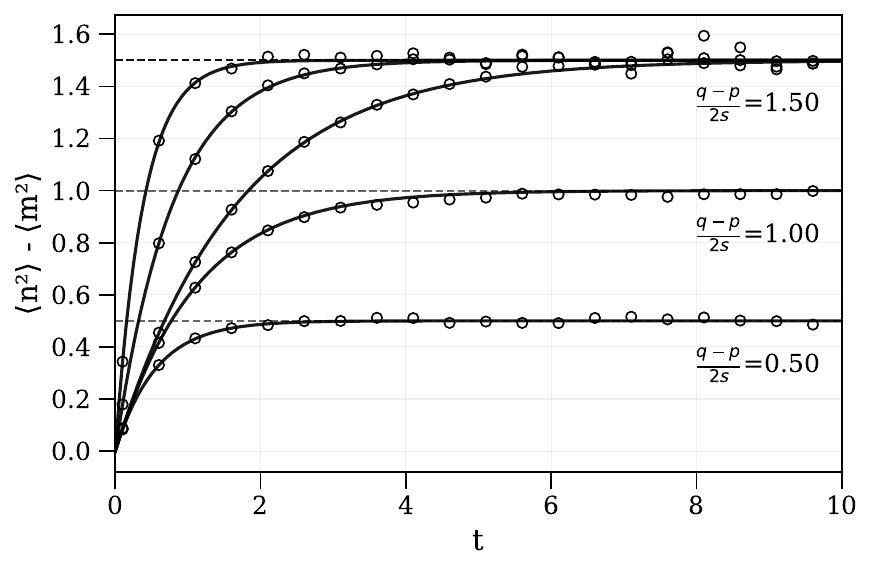}
  \caption{Asymptotic difference between the second moments $\langle n^2 \rangle_v - \langle m^2 \rangle_v$ versus $s^{-1}$ and $q>p$ showing convergence to $(q-p)/(2s)$ for three sets of parameters: $(q,p,s)$=$\{0.1,4.0,1.95\},\{0.1,0.5,0.2\},\{0.1,1.0,0.45\}$ and $\{0.1,2.0,0.95\}$. Lines: theory; points: simulations. }
  \label{fig:var_diff}
\end{figure}

A key finding of our analysis is that the swap process does not merely average the mobilities. Indeed, the difference between the second moments computed in both channels $\langle n^2 \rangle_v - \langle m^2 \rangle_v = \frac{q-p}{2s}(1 - e^{-2st})$ approaches a constant value $\frac{q-p}{2s}$ as $t \to \infty$. This asymptotic behavior reveals that swap processes effectively freeze mobility mismatches into persistent anisotropy (see Fig. \ref{fig:var_diff} for a comparison with numerical simulations). This difference serves as a quantitative measure to characterize transport efficiency in motor teams (e.g., cargo dispersion) and quantify risk asymmetry in financial markets (e.g., volatility imbalance). It also provides a mechanistic explanation for why certain synapses emerge as dominant memory storage sites—engram cells—despite operating within a shared biochemical environment.

The time-dependent diffusion exponent $\alpha(t) = d \log \langle n^2(t) \rangle_v / d \log t$ can be computed directly from eq. (\ref{eq:variances}). For $n_0=m_0$, it exhibits a crossover:
\begin{equation}\label{eq.diff_coef_n00}
	\alpha_{n,m}(t) = \frac{1 \pm \frac{q-p}{q+p} e^{-2st}}{1 \pm \frac{q-p}{q+p} \frac{1-e^{-2st}}{2st}}
\end{equation}
where $+$ ($-$) corresponds to the particle with higher (lower) transition rate.
At intermediate timescales, the faster particle exhibits transient subdiffusion ($\alpha \leq 1$) while the slower particle displays superdiffusive behavior ($\alpha \geq 1$). Both ultimately converge to normal diffusion with an effective diffusion coefficient $(q+p)/2$ at long times. This dynamical crossover in the diffusion mimics the enhanced dispersal of kinesin-driven vesicles during motor switching events \cite{Caspi00} and the ``momentum bursts" of high-volatility assets during liquidity crises \cite{Gabaix06}.

At this point it is important to notice that our model admits an alternative interpretation that connects it with the broader class of switching diffusion processes. Rather than two particles exchanging spatial positions, our formulation can be viewed as describing two particles that stochastically interchange their diffusive modes characterized by rates $p$ and $q$, where the switching itself occurs at rate $s$. This perspective aligns our work with recent advances in intermittent active motion \cite{Datta24} and the foundational double-diffusivity theory developed in \cite{Showalter04}, to models transport in heterogeneous media. However, our approach differs fundamentally from previous switching models \cite{Santra24, Chatto23}: whereas classical treatments consider particles independently transitioning between different diffusive states, our mechanism enforces a correlated exchange where the diffusion rates are swapped between two coordinated agents introducing a form of memory and reciprocity not captured in other switching models.


In conclusion, we have introduced an exactly solvable model of two random walkers with heterogeneous mobilities and stochastic position exchange. The joint distribution function $P_{n,m}(t)$ captures dynamical crossovers from initial independent diffusion to swap-dominated synchronization, where both particles diffuse at identical rate $\frac{q+p}{2}$, but swapping positions, and to a regime where the mean position of particles in both channels is the same. Notably, during the first crossover, the slower particle exhibits transient superdiffusion while the faster one becomes subdiffusive. Our analysis also reveals  the existence of a persistent heterogeneity, defined by the difference $\langle n^2\rangle - \langle m^2 \rangle  = \frac{q-p}{2s}$, that quantifies how swap-mediated correlations freeze mobility mismatch into spatial anisotropy.  Although our model ignores complexities like load-dependent motor kinetics, directional motion, distance dependence of the swapping rate, or trader adaptation, it successfully isolates the universal role played by stochastic exchange in mediating correlations between heterogeneous agents. The analytical solution presented here constitutes a valuable benchmark for extending the model to these richer scenarios.

This work was supported by the Marie Skłodowska Curie Action (MSCA) Staff Exchange project “SIMBAD” (REA Grant Agreement n. 101131463).

\bibliography{ref_swapping}

\newpage
\onecolumngrid

\section{Supplementary Materials}
We consider two particles undergoing independent one-dimensional random walks on parallel lines. Each particle moves to the left or right with rates $q$ and $p$, respectively, which may depend on the line they occupy. Additionally, the particles can exchange their positions across the two lines at a constant rate $s$. Specifically, if the particles are at positions $n$ (line 1) and $m$ (line 2) at time $t$, they may switch positions, resulting in the configuration $m$ (line 1) and $n$ (line 2), with rate $s$.
\subsection{Master equation of the process.}
We aim to compute the time evolution of the joint probability $P_{n,m}(t)$, which gives the probability of finding one particle at position $n$ on line 1 and the other at position $m$ on line 2 at time $t$. The marginal distributions are defined as:
\begin{equation}
	\sum_m P_{n,m}(t) = P_n(t), \qquad \sum_n P_{n,m}(t) = P_m(t),
\end{equation}
where $P_n(t)$ and $P_m(t)$ denote the probabilities of finding a particle at position $n$ or $m$ on their respective lines. The joint probability is normalized so that
\begin{equation}
	\sum_{n,m} P_{n,m}(t) = 1.
\end{equation}
In this case the master equation should take the form:
\begin{equation}
  \dot{P}_{n,m} = \sum_{\alpha,\beta} W_{(\alpha \beta) (n m)} P_{\alpha,\beta} - \sum_{\alpha,\beta}
  W_{(n m) (\alpha \beta) } P_{n, m}
  \label{eq:ME2p}
\end{equation}
where $W_{(\alpha \beta) (n m)}$ is a transition rate from having particles in the positions $\alpha$ and  $\beta$ in the first and second lines, to states $n$ and $m$ also in the first and second lines. We define the transition rates as:
\begin{equation}
  W_{(\alpha \beta) (n m)} = \frac{q}{2} \Big[  \delta_{\alpha,n-1} + \delta_{\alpha,n+1} \Big] \delta_{\beta,m} + \frac{p}{2} \Big[  \delta_{\beta,m-1} + \delta_{\beta,m+1} \Big] \delta_{\alpha,n} + s \delta_{\beta,n} \delta_{\alpha,m}  + (1-q-p-s) \delta_{\alpha,n} \delta_{\beta,m}
\label{eq:TranRates2p}
\end{equation}
where the first two terms describe single-particle motion while the other remains stationary and the third term, proportional to $s$, accounts for the exchange of particle positions between the two lines. The last term represents the probability of both particles staying at their current positions.
Summing over $\alpha$ and $\beta$ yields:
\begin{eqnarray}
\sum_{\alpha,\beta} W_{(\alpha \beta) (n m)} P_{\alpha,\beta} &=& \frac{q}{2} \Big[ P_{n-1, m} + P_{n+1,m} \Big] + \frac{p}{2} \Big[ P_{n, m-1} + P_{n,m+1} \Big] + s P_{m,n} + (1-q-p-s)P_{n,m}
\end{eqnarray}
and similarly:
\begin{equation}
  \left(\sum_{\alpha,\beta} W_{(n m) (\alpha \beta)}\right) P_{n,m} = \left[ \frac{q}{2}+ \frac{q}{2}+\frac{p}{2}+\frac{p}{2} + s + ( 1 - q - p -s)\right] P_{n,m}  = P_{n,m}
\end{equation}
Collecting terms, we arrive at the following master equation for the joint probability:	
\begin{equation}
  \dot{P}_{n,m} = \frac{q}{2}\Big[ P_{n-1, m} + P_{n+1,m} \Big] + \frac{p}{2} \Big[ P_{n, m-1} + P_{n,m+1} \Big] - (q + p + s) P_{n,m} + s P_{m,n} 
  \label{eq:RWalk2p}
\end{equation}
\subsection{Closed-Form Solution of the Master Equation in Laplace–Fourier Space}
We now apply a Fourier transform in space and a Laplace transform in time to Eq.~(\ref{eq:RWalk2p}) to obtain an analytical solution.
 Defining: $G(k_x,k_y,t)= \sum_{n,m} e^{i k_x n + i k_y m} P_{n,m}(t)$ and substituting this  into (\ref{eq:RWalk2p}) we get the following differential equation:
\begin{eqnarray}
  \dot{G}(k_x,k_y,t) &=& \frac{q}{2} \Big[ e^{i k_x} +  e^{-i k_x} \Big] G +
  \frac{p}{2} \Big[ e^{i k_y} +  e^{-i k_y} \Big] G - (q+p+s) G + s \sum_{n,m} e^{i k_x n + i k_y m} P_{m,n} \nonumber \\
  &=&  \Big\{ q \Big[ \cos(k_x)-1\Big] + p \Big[ \cos(k_y)-1\Big]  - s \Big\} G(k_x,k_y,t) + sG(k_y,k_x,t) 
  \label{eq:Equation2D}.
  \end{eqnarray}

For notational simplicity, we define
\[
w(k_x, k_y) = q \left[ \cos(k_x) - 1 \right] + p \left[ \cos(k_y) - 1 \right] - s,
\]
as well as \( \vec{k} = (k_x, k_y) \), \( \underline{\vec{k}} = (k_y, k_x) \), and write \( w\left(\vec{k}\right) \equiv w(k_x, k_y) \).  For simplicity, we suppress the vector notation in this Supplementary Material. However, the meaning remains consistent with the notation used in the main text.
With this notation, Eq.~(\ref{eq:Equation2D}) can be rewritten as:
\begin{equation}
  \dot{G}\left(\vec{k},t\right) =  w\left(\vec{k}\right) G\left(\vec{k},t\right) + s \int d\vec{k}'  G\left(\vec{k}',t\right) \delta\left(\vec{k}',\vec{\underline{k}}\right)
  \label{eq:Equation2D-b}
  \end{equation}
Transforming to Laplace, this leads to:
\begin{equation}
  r \hat{G}\left(\vec{k},r\right) - \hat{G}\left(\vec{k},0\right) =  w\left(\vec{k}\right) \hat{G}\left(\vec{k},r\right) + s \int d\vec{k}'  G\left(\vec{k}',r\right) \delta\left(\vec{k}',\vec{\underline{k}}\right)
  \label{eq:Equation2D-Laplace_1}
  \end{equation}
that can be rewritten as:
\begin{equation}
  \hat{G}\left(\vec{k},r\right)= \frac{\hat{G}\left(\vec{k},0\right)}{r - w\left(\vec{k}\right)}  + \frac{s}{r - w\left(\vec{k}\right)} \int d\vec{k}'  G\left(\vec{k}',r\right) \delta\left(\vec{k}',\vec{\underline{k}}\right)
  \label{eq:Equation2D-Laplace_2}
  \end{equation}

Fredholm equation (\ref{eq:Equation2D-Laplace_2}) can be solved through the method of successive iterations. Solving Eq.~(\ref{eq:Equation2D-Laplace_2}) via successive substitution yields the following series:
\begin{eqnarray}
  \hat{G}\left(\vec{k},r\right) & = & \frac{\hat{G}\left(\vec{k},0\right)}{r - w\left(\vec{k}\right)} +
  \frac{s\hat{G}\left(\underline{\vec{k}},0\right)}{\left(r-w\left(\vec{\underline{k}}\right)\right)\left(r - w\left(\vec{k}\right)\right)} +
  \frac{s^2\hat{G}\left(\vec{k},0\right)}{\left(r-  w\left(\vec{\underline{k}}\right)\right) \left(r-w\left(\vec{k}\right)\right)^2} + \frac{s^3\hat{G}\left(\underline{\vec{k}},0\right)}{\left(r-w\left(\vec{\underline{k}}\right)\right)^2 \left(r-w\left(\vec{k}\right)\right)^2 } + \dots   \nonumber\\
  &=&  \frac{\hat{G}\left(\vec{k},0\right)}{r - w\left(\vec{k}\right)} +
  \frac{s\hat{G}\left(\underline{\vec{k}},0\right)}{\left(r-w\left(\vec{\underline{k}}\right)\right)\left(r - w\left(\vec{k}\right)\right)} +
  \frac{s^2\hat{G}(k,r)}{\left(r-w\left(\vec{\underline{k}}\right)\right) \left(r-w\left(\vec{k}\right)\right)}
\end{eqnarray}
Noticing that \( \hat{G}\left(\vec{k}, r\right) \) reappears on the right-hand side of the series, we identify this recurrence and solve the resulting algebraic equation to obtain the closed-form expression:
\begin{equation}
\hat{G}\left(\vec{k},r\right) =  \frac{\hat{G}\left(\vec{k},0\right) \left(r-w\left(\vec{\underline{k}}\right)\right)+s\hat{G}\left(\underline{\vec{k}},0\right) }{\left(r-w\left(\vec{k}\right)\right)\left(r-w\left(\vec{\underline{k}}\right)\right)-s^2}
\label{eq:sol2D-Laplace}
\end{equation}
This expression provides the exact Laplace–Fourier domain solution for the joint probability distribution.

\subsection{Closed Solution for the Probability Distribution}

\subsubsection*{Inverse Laplace Transform: Recovering Time}

We begin by rewriting Eq.~(\ref{eq:sol2D-Laplace}) in the form:
\begin{equation}
	\hat{G}\left(\vec{k}, r\right) = \frac{ \hat{G}\left(\vec{k},0\right)(r - w\left(\vec{\underline{k}}\right)) + s \hat{G}\left(\underline{\vec{k}}, 0\right) }{(r - r_1)(r - r_2)},
	\label{eq:sol2D-Laplace-2}
\end{equation}
where the poles are given by
\[
r_{1,2} = \frac{1}{2} \left[ w\left(\vec{k}\right) + w\left(\vec{\underline{k}}\right) \pm \sqrt{ \left(w\left(\vec{k}\right) - w\left(\vec{\underline{k}}\right)\right)^2 + 4s^2 } \right].
\]
For notational simplicity, we define $\Delta := w\left(\vec{k}\right) - w\left(\vec{\underline{k}}\right).$
Computing the inverse Laplace transform (Mellin integral), we find:
\begin{equation}
	G\left(\vec{k}, t\right) = 
	\frac{
		s G\left(\underline{\vec{k}}, 0\right) + \frac{G\left(\vec{k}, 0\right)}{2} \left( \Delta + \sqrt{\Delta^2 + 4s^2} \right)
	}{
		\sqrt{\Delta^2 + 4s^2}
	} \, \mathrm{e}^{r_1 t}
	-
	\frac{
		s G\left(\underline{\vec{k}}, 0\right) + \frac{G\left(\vec{k}, 0\right)}{2} \left( \Delta - \sqrt{\Delta^2 + 4s^2} \right)
	}{
		\sqrt{\Delta^2 + 4s^2}
	} \, \mathrm{e}^{r_2 t}.
	\label{eq:G(k,t)_}
\end{equation}
In the limit \( s \to 0 \), this expression reduces to the product of two independent random walkers.

\subsubsection*{Inverse Fourier Transform: Recovering Space}
To recover the probability distribution \( P_{n,m}(t) \), we compute the inverse Fourier transform of Eq.~(\ref{eq:G(k,t)_}). Note that:
\[\Delta = 
w\left(\vec{k}\right) - w\left(\vec{\underline{k}}\right) = (q - p)(\cos k_x - \cos k_y), \quad
w\left(\vec{k}\right) + w\left(\vec{\underline{k}}\right) = (q + p)(\cos k_x + \cos k_y - 2) - 2s.
\]
Substituting into Eq.~(\ref{eq:G(k,t)_}) and taking the inverse transform, we write:
\begin{align}
	P_{n,m}(t) &= \frac{e^{-(q + p + s)t}}{2} \frac{1}{(2\pi)^2} \int dk_x dk_y \, e^{-i k_x (n - n_0) - i k_y (m - m_0)} e^{\frac{t}{2}(q + p)(\cos k_x + \cos k_y)} \nonumber \\
	&\quad \times \Bigg[
	e^{\frac{t}{2} \sqrt{\Delta^2 + 4s^2}} \left(1 + \frac{\Delta + 2s e^{i(k_x (m_0 - n_0) + k_y (n_0 - m_0))}}{\sqrt{\Delta^2 + 4s^2}} \right) +
	e^{-\frac{t}{2} \sqrt{\Delta^2 + 4s^2}} \left(1 - \frac{\Delta + 2s e^{i(k_x (m_0 - n_0) + k_y (n_0 - m_0))}}{\sqrt{\Delta^2 + 4s^2}} \right)
	\Bigg].
\end{align}
To facilitate the inverse Fourier transform, we introduce an auxiliary integral representation of the Dirac delta function:
$
\delta(\psi - \Delta) = \frac{1}{2\pi} \int d\hat{\psi} \, e^{i\hat{\psi}(\psi - \Delta)}.
$
This technique allows us to isolate the dependence on the variable $\Delta$ and manage the nonlinear structure of the integrand more effectively during the calculation.
Substituting this into the integral and integrating over \( \psi \), we find:
\begin{align}
	P_{n,m}(t) &= \frac{e^{-(q + p + s)t}}{2\pi} \int d\psi \, d\hat{\psi} \, e^{i\psi \hat{\psi}} \nonumber  \left[
	\cosh\left(\frac{t}{2} \sqrt{\psi^2 + 4s^2} \right)
	+ \frac{\psi}{\sqrt{\psi^2 + 4s^2}} \sinh\left(\frac{t}{2} \sqrt{\psi^2 + 4s^2} \right)
	\right] \nonumber \\
	&\quad \times I_{n-n_0}\left( \frac{q+p}{2}t - i\hat{\psi}(q - p) \right)
	I_{m-m_0}\left( \frac{q+p}{2}t + i\hat{\psi}(q - p) \right) \nonumber \\
	&\quad + \frac{e^{-(q + p + s)t}}{2\pi} \int d\psi \, d\hat{\psi} \, e^{i\psi \hat{\psi}} \cdot
	\left[
	\frac{2s}{\sqrt{\psi^2 + 4s^2}} \sinh\left(\frac{t}{2} \sqrt{\psi^2 + 4s^2} \right)
	\right] \nonumber \\
	&\quad \times I_{n - m_0}\left( \frac{q + p}{2}t - i\hat{\psi}(q - p) \right)
	I_{m - n_0}\left( \frac{q + p}{2}t + i\hat{\psi}(q - p) \right),
	\label{eq:pre_final_full}
\end{align}
where \( I_n(\cdot) \) is the modified Bessel function of the first kind.

To handle the integrals over \( \psi \), we regularize them using a Gaussian cutoff:
\begin{align}
	\mathcal{I}_1 &= \lim_{\epsilon \to 0} \int d\psi \, e^{i\psi \hat{\psi} - \epsilon^2 \psi^2} \cosh\left(\frac{t}{2} \sqrt{\psi^2 + 4s^2} \right), \\
	\mathcal{I}_2 &= \lim_{\epsilon \to 0} \int d\psi \, e^{i\psi \hat{\psi} - \epsilon^2 \psi^2}
	\left[ \frac{\psi + 2s}{\sqrt{\psi^2 + 4s^2}} \sinh\left(\frac{t}{2} \sqrt{\psi^2 + 4s^2} \right) \right].
\end{align}

\bigskip

 \subsubsection{Computing $\mathcal{I}_1$}

We begin by working with the series expansion of $\cosh$, which allows us to write:
$$
\mathcal{I}_1 = \lim_{\epsilon \rightarrow 0} \sum_{k=0}^\infty \left(\frac{t}{2}\right)^{2k} \frac{1}{(2k)!}
\int d\psi \, e^{i \psi \hat{\psi} - \epsilon^2 \psi^2} \left(\psi^2 + 4s^2\right)^k.
$$
Expanding the binomial and reorganizing terms, we obtain:
$$
\mathcal{I}_1 = \lim_{\epsilon \rightarrow 0} \sum_{k=0}^\infty (st)^{2k} \frac{1}{(2k)!}
\sum_{l=0}^k \binom{k}{l} (4s^2)^{-l} \int d\psi \, e^{i \psi \hat{\psi} - \epsilon^2 \psi^2} \psi^{2l}.
$$
The integral can be evaluated using the identity (from the table of integral transforms, Eq. (3.20)):
$$
\int d\psi \, e^{i \psi \hat{\psi} - \epsilon^2 \psi^2} \psi^{2l}
= 2 \int_0^\infty d\psi \, e^{-\epsilon^2 \psi^2} \cos(\hat{\psi} \psi) \psi^{2l}
= 2 (-1)^l \sqrt{\pi} 2^{-l-1} \epsilon^{-2l-1} e^{-\frac{\hat{\psi}^2}{4 \epsilon^2}} H_{e_{2l}}\left( \frac{\hat{\psi}}{\sqrt{2} \epsilon} \right).
$$
Substituting back into the sum and rearranging:
$$
\mathcal{I}_1 = \lim_{\epsilon \to 0} \frac{\sqrt{\pi}}{\epsilon} e^{-\frac{\hat{\psi}^2}{4 \epsilon^2}} \sum_{l=0}^\infty (-1)^l \left( \frac{1}{2s\sqrt{2} \epsilon} \right)^{2l} H_{e_{2l}}\left( \frac{\hat{\psi}}{\sqrt{2} \epsilon} \right)
\sum_{k=l}^\infty \binom{k}{l} (st)^{2k} \frac{1}{(2k)!}.
$$
The inner sum evaluates to:
$$
\sum_{k=l}^\infty \binom{k}{l} (st)^{2k} \frac{1}{(2k)!} = \frac{\sqrt{\pi} \, 2^{-l - 1/2} (st)^{l + 1/2}}{\Gamma(l + 1)} I_{l - 1/2}(st).
$$
Substituting this result yields:
$$
\mathcal{I}_1 = \lim_{\epsilon \to 0} \sqrt{\frac{st}{2}} \frac{\pi}{\epsilon} e^{-\frac{\hat{\psi}^2}{4 \epsilon^2}} \sum_{l=0}^\infty (-1)^l \frac{(st)^l}{l!} \left( \frac{1}{4s \epsilon} \right)^{2l} H_{e_{2l}}\left( \frac{\hat{\psi}}{\sqrt{2} \epsilon} \right) I_{l - 1/2}(st).
$$
Using the series form of the Bessel function $I_{l - 1/2}$:
$$
I_{l - 1/2}(st) = \left( \frac{st}{2} \right)^{l - 1/2} \sum_{k=0}^\infty \frac{1}{k! \Gamma(k + l + 1/2)} \left( \frac{st}{2} \right)^{2k},
$$
we obtain:
$$
\mathcal{I}_1 = \lim_{\epsilon \to 0} \frac{\pi}{\epsilon} e^{-\frac{\hat{\psi}^2}{4 \epsilon^2}} \sum_{l=0}^\infty (-1)^l \frac{1}{l!} \left( \frac{t}{4 \sqrt{2} \epsilon} \right)^{2l} H_{e_{2l}}\left( \frac{\hat{\psi}}{\sqrt{2} \epsilon} \right)
\left[ \frac{1}{\Gamma(l + 1/2)} + \sum_{k=1}^\infty \frac{1}{k! \Gamma(k + l + 1/2)} \left( \frac{st}{2} \right)^{2k} \right].
$$
Now, rewriting the reciprocal Gamma term using:
$$
\frac{1}{\Gamma(k + l + 1/2)} = \frac{1}{\Gamma(l + 1/2) \Gamma(k)} \int_0^1 dz\, z^{l - 1/2} (1 - z)^{k - 1},
$$
we substitute into the sum:
$$
\mathcal{I}_1 = \lim_{\epsilon \to 0} \frac{\pi}{\epsilon} e^{-\frac{\hat{\psi}^2}{4 \epsilon^2}} \sum_{l=0}^\infty \frac{(-1)^l}{l! \, \Gamma(l + 1/2)} \left( \frac{t}{4 \sqrt{2} \epsilon} \right)^{2l} H_{e_{2l}}\left( \frac{\hat{\psi}}{\sqrt{2} \epsilon} \right)
\left[ 1 + \int_0^1 dz\, z^{l - 1/2} (1 - z)^{-1} \sum_{k=1}^\infty \frac{1}{k! \Gamma(k)} \left( \frac{st}{2} \right)^{2k} (1 - z)^k \right].
$$
Evaluating the inner sum:
\begin{equation}
S_1 = \sum_{k=1}^\infty \frac{1}{k! \Gamma(k)} \left( \left( \frac{st}{2} \right)^2 (1 - z) \right)^k = \frac{st}{2} \sqrt{1 - z} \, I_1\left( st \sqrt{1 - z} \right),\label{eq:inner_sum}
\end{equation}
we find:
$$
\mathcal{I}_1 = \lim_{\epsilon \to 0} \frac{\pi}{\epsilon} e^{-\frac{\hat{\psi}^2}{4 \epsilon^2}} \sum_{l=0}^\infty \frac{(-1)^l}{l! \, \Gamma(l + 1/2)} \left( \frac{t}{4 \sqrt{2} \epsilon} \right)^{2l} H_{e_{2l}}\left( \frac{\hat{\psi}}{\sqrt{2} \epsilon} \right)
\left[ 1 + \frac{st}{2} \int_0^1 dz\, \frac{z^{l - 1/2}}{\sqrt{1 - z}} I_1(st \sqrt{1 - z}) \right].
$$
We now simplify the sum over $l$ using $\Gamma(l + 1/2) = \sqrt{\pi} \, \frac{(2l)!}{4^l l!}$, and the identity for Hermite polynomials:
$$
H_{e_n}(x) = 2^{-n/2} H_n\left( \frac{x}{\sqrt{2}} \right),
$$
leading to:
$$
\sum_{l=0}^\infty \frac{(-1)^l}{l! \, \Gamma(l + 1/2)} \left( \frac{t}{4 \sqrt{2} \epsilon} \right)^{2l} H_{e_{2l}}\left( \frac{\hat{\psi}}{\sqrt{2} \epsilon} \right) =
\frac{1}{\sqrt{\pi}} \sum_{l=0}^\infty \frac{(-1)^l}{(2l)!} \left( \frac{t}{4 \epsilon} \right)^{2l} H_{2l}\left( \frac{\hat{\psi}}{2 \epsilon} \right).
$$
Using the generating function identity:
$$
\sum_{l=0}^\infty \frac{(-1)^l}{(2l)!} y^{2l} H_{2l}(x) = e^{y^2} \cos(2 x y),
$$
we find:
$$
S_2 = \frac{1}{\sqrt{\pi}} e^{\frac{t^2}{16 \epsilon^2}} \cos\left( \frac{\hat{\psi} t}{4 \epsilon^2} \right).
$$
Putting all pieces together:
$$
\mathcal{I}_1 = \lim_{\epsilon \to 0} \frac{\pi}{\epsilon} e^{-\frac{\hat{\psi}^2}{4 \epsilon^2}} \left[
\frac{1}{\sqrt{\pi}} e^{\frac{t^2}{16 \epsilon^2}} \cos\left( \frac{\hat{\psi} t}{4 \epsilon^2} \right)
+ \frac{st}{2} \int_0^1 dz\, \frac{1}{\sqrt{z(1 - z)}} I_1(st \sqrt{1 - z}) \frac{1}{\sqrt{\pi}} e^{\frac{t^2 z}{16 \epsilon^2}} \cos\left( \frac{\hat{\psi} t}{4 \epsilon^2} \sqrt{z} \right)
\right].
$$
Finally, using the identity:
$$
\lim_{\epsilon \to 0} \frac{1}{\sqrt{2\pi} \epsilon} e^{- \frac{x^2}{2 \epsilon^2}} = \delta(x),
$$
we arrive at the final expression:
$$
	\mathcal{I}_1 = \pi \left[ \delta\left(\hat{\psi} - i \frac{t}{2} \right) + \delta\left(\hat{\psi} + i \frac{t}{2} \right)
	+ \frac{st}{2} \int_0^1 dz\, \frac{I_1(st \sqrt{1 - z})}{\sqrt{z(1 - z)}}  \left( \delta\left(\hat{\psi} - i \frac{t}{2} \sqrt{z} \right) + \delta\left(\hat{\psi} + i \frac{t}{2} \sqrt{z} \right) \right)
	\right].
$$

\subsubsection{Computing $\mathcal{I}_2$}

Proceeding in a similar way as in the computation for $\cosh$, we begin with:
\begin{eqnarray}\label{ec:43}
	\mathcal{I}_2 &=& \lim_{\epsilon \rightarrow 0} \sum_{k=0}^\infty \left(\frac{t}{2}\right)^{2k+1}
	\frac{1}{ (2k+1)!} \int d\psi \, e^{i \psi \hat{\psi} - \epsilon^2 \psi^2}  \left(\psi+2s\right) \left(\psi^2+4s^2\right)^{k} \nonumber \\
	&=& \lim_{\epsilon \rightarrow 0} \sum_{k=0}^\infty (st)^{2k+1} \frac{1}{ (2k+1)!}
	\sum_{l=0}^k \binom{k}{l} (4s^2)^{-l} \int d\psi \, e^{i \psi \hat{\psi}- \epsilon^2 \psi^2} \psi^{2l} \left(1+\frac{\psi}{2s}\right)
\end{eqnarray}
At this point, we recall the integral identities:
\begin{align*}
	\int d\psi \, e^{i \psi \hat{\psi}- \epsilon^2 \psi^2} \psi^{2l} &= 2 \int_0^\infty d\psi \, e^{- \epsilon^2 \psi^2} \cos(\hat{\psi} \psi) \psi^{2l} = 2 (-1)^l \sqrt{\pi} 2^{-l-1} \epsilon^{-2l-1} e^{- \frac{ \hat{\psi}^2}{4\epsilon^2}} H_{e_{2l}} \left( \frac{\hat{\psi}}{\sqrt{2}\epsilon} \right), \\
	\int d\psi \, e^{i \psi \hat{\psi}- \epsilon^2 \psi^2} \psi^{2l+1} &= 2 i \int_0^\infty d\psi \, e^{- \epsilon^2 \psi^2} \sin(\hat{\psi} \psi) \psi^{2l+1} = 2i (-1)^l \sqrt{\pi} 2^{-l - 3/2} \epsilon^{-2l-2} e^{- \frac{ \hat{\psi}^2}{4\epsilon^2}} H_{e_{2l+1}} \left( \frac{\hat{\psi}}{\sqrt{2}\epsilon} \right).
\end{align*}
Substituting back into (\ref{ec:43}), we obtain:
\begin{eqnarray}
	\mathcal{I}_2
	&=& \lim_{\epsilon \rightarrow 0} \frac{\sqrt{\pi}}{\epsilon} e^{- \frac{ \hat{\psi}^2}{4\epsilon^2}} \sum_{k=0}^\infty \frac{ (st)^{2k+1}}{ (2k+1)!}
	\sum_{l=0}^k \binom{k}{l} (-1)^l (4s^2)^{-l} \left[
	2^{-l} \epsilon^{-2l}  H_{e_{2l}} \left( \frac{\hat{\psi}}{\sqrt{2}\epsilon} \right) + \frac{i}{2s} 2^{-l - 1/2} \epsilon^{-2l - 1}  H_{e_{2l+1}} \left( \frac{\hat{\psi}}{\sqrt{2}\epsilon} \right)
	\right] \nonumber \\
	&=& \lim_{\epsilon \rightarrow 0} \frac{\sqrt{\pi}}{\epsilon} e^{- \frac{ \hat{\psi}^2}{4\epsilon^2}}
	\sum_{l=0}^\infty (-1)^l \left[
	\left(\frac{1}{ 2 s \epsilon \sqrt{2}}\right)^{2l} H_{e_{2l}} \left( \frac{\hat{\psi}}{\sqrt{2}\epsilon} \right) + i \left(\frac{1}{2s \epsilon \sqrt{2}}\right)^{2l+1} H_{e_{2l+1}} \left( \frac{\hat{\psi}}{\sqrt{2}\epsilon} \right) \right]
	\sum_{k=l}^\infty \binom{k}{l} \frac{ (st)^{2k+1}}{ (2k+1)!}
\end{eqnarray}
Now, using:
\[
\sum_{k=l}^\infty \binom{k}{l} \frac{(st)^{2k+1}}{(2k+1)!} = (2st)^{l+1/2} \frac{\Gamma(l+3/2)}{\Gamma(2l+2)} I_{l+1/2}(st),
\]
we find:
\begin{equation}
	\mathcal{I}_2 =
	\lim_{\epsilon \rightarrow 0} \frac{\sqrt{\pi}}{\epsilon} e^{- \frac{ \hat{\psi}^2}{4\epsilon^2}}
	\sum_{l=0}^\infty (-1)^l \left( \frac{1}{ 2 s \epsilon \sqrt{2}} \right)^{2l} \left[
	H_{e_{2l}} \left( \frac{\hat{\psi}}{\sqrt{2}\epsilon} \right) + \left( \frac{i}{2s \epsilon \sqrt{2}} \right) H_{e_{2l+1}} \left( \frac{\hat{\psi}}{\sqrt{2}\epsilon} \right) \right]
	(2st)^{l+1/2} \frac{\Gamma\left(l+3/2\right)}{\Gamma(2l+2)} I_{l+1/2}(st)
\end{equation}
Using the series representation:
\[
I_{l+1/2}(st) = \left(\frac{st}{2}\right)^{l+1/2} \sum_{k=0}^\infty \frac{1}{k! \Gamma(k + l + 3/2)} \left( \frac{st}{2} \right)^{2k},
\]
we substitute and separate into two parts:
\begin{eqnarray}
	\mathcal{I}_2 &=& \lim_{\epsilon \rightarrow 0} \frac{\sqrt{\pi}}{\epsilon} e^{- \frac{ \hat{\psi}^2}{4\epsilon^2}}
	\sum_{l=0}^\infty (-1)^l \left[
	st \left( \frac{t}{2 \epsilon \sqrt{2}} \right)^{2l} H_{e_{2l}} \left( \frac{\hat{\psi}}{\sqrt{2}\epsilon} \right) + i \left( \frac{t}{2 \epsilon \sqrt{2}} \right)^{2l+1} H_{e_{2l+1}} \left( \frac{\hat{\psi}}{\sqrt{2}\epsilon} \right) \right] \frac{\Gamma(l+3/2)}{\Gamma(2l+2)} \sum_{k=0}^\infty \frac{\left( \frac{st}{2} \right)^{2k}}{k! \Gamma(k + l + 3/2)}
\end{eqnarray}
We now treat both contributions separately. First:

$$
\begin{aligned}
	\mathcal{I}_{2a} &= \lim_{\epsilon \to 0} \frac{\sqrt{\pi}}{\epsilon} e^{ -\frac{1}{4} \frac{\hat{\psi}^2}{\epsilon^2} }
	i \sum_{l=0}^\infty (-1)^l \left[ \left(\frac{t}{2 \epsilon \sqrt{2}} \right)^{2l+1} H_{e_{2l+1}} \left( \frac{\hat{\psi}}{ \sqrt{2} \epsilon } \right) \right] \frac{ \Gamma(l+3/2) }{ \Gamma(2l+2) } \sum_{k=0}^\infty \frac{1}{k! \Gamma(k+l+3/2)} \left( \frac{st}{2} \right)^{2k}
\end{aligned}
$$
Now use the identity:
$$
\frac{1}{ \Gamma(k+l+3/2) } = \frac{ \int_0^1 dz\, z^{l+1/2} (1-z)^{k-1} }{ \Gamma(l+3/2) \Gamma(k) }
$$
So:
$$
\begin{aligned}
	\mathcal{I}_{2a} &= \lim_{\epsilon \to 0} \frac{\sqrt{\pi}}{\epsilon} e^{ -\frac{1}{4} \frac{\hat{\psi}^2}{\epsilon^2} } i \sum_{l=0}^\infty \frac{ (-1)^l }{ \Gamma(2l+2) } \left[ \left( \frac{t}{2 \epsilon \sqrt{2}} \right)^{2l+1} H_{e_{2l+1}} \left( \frac{\hat{\psi}}{ \sqrt{2} \epsilon } \right) \right] \left[ 1 + \int_0^1 dz\, z^{l+1/2} \sum_{k=1}^\infty \frac{1}{k! \Gamma(k)} \left( \frac{st}{2} \right)^{2k} (1-z)^{k-1} \right]
\end{aligned}
$$
Recognizing the identity (\ref{eq:inner_sum}), we get:
$$
\begin{aligned}
	\mathcal{I}_{2a} &= \lim_{\epsilon \to 0} \frac{\sqrt{\pi}}{\epsilon} e^{ -\frac{1}{4} \frac{\hat{\psi}^2}{\epsilon^2} } i \sum_{l=0}^\infty \frac{ (-1)^l }{ \Gamma(2l+2) } \left[ \left( \frac{t}{2 \epsilon \sqrt{2}} \right)^{2l+1} H_{e_{2l+1}} \left( \frac{\hat{\psi}}{ \sqrt{2} \epsilon } \right) \right] \left[ 1 + \frac{st}{2} \int_0^1 dz\, z^{l+1/2} \frac{1}{ \sqrt{1-z} } I_1(st \sqrt{1-z}) \right]
\end{aligned}
$$
Now focus on the sum over $l$:
$$
\begin{aligned}
	S_2 &= \sum_{l=0}^\infty \frac{ (-1)^l }{ \Gamma(2l+2) } \left( \frac{t}{2 \epsilon \sqrt{2}} \right)^{2l+1} H_{e_{2l+1}} \left( \frac{ \hat{\psi} }{ \sqrt{2} \epsilon } \right) = \sum_{l=0}^\infty \frac{ (-1)^l }{ (2l+1)! } \left( \frac{t}{4\epsilon} \right)^{2l+1} H_{2l+1} \left( \frac{ \hat{\psi} }{2\epsilon} \right) = e^{ \frac{t^2}{16 \epsilon^2} } \sin\left( \frac{t \hat{\psi}}{4 \epsilon^2} \right)
\end{aligned}
$$
Then the contribution becomes:
$$
\begin{aligned}
	\mathcal{I}_{2a} &= \lim_{\epsilon \to 0} \frac{\sqrt{\pi}}{2\epsilon} e^{ -\frac{1}{4} \frac{\hat{\psi}^2}{\epsilon^2} } \left[ e^{ \frac{t^2}{16 \epsilon^2} } \left( e^{ i \frac{t \hat{\psi}}{4 \epsilon^2} } - e^{ -i \frac{t \hat{\psi}}{4 \epsilon^2} } \right) + \frac{st}{2} \int_0^1 dz\, e^{ \frac{t^2 z}{16 \epsilon^2} } \frac{1}{ \sqrt{1-z} } I_1(st \sqrt{1-z}) \left( e^{ i \frac{t \hat{\psi} \sqrt{z} }{4 \epsilon^2} } - e^{ -i \frac{t \hat{\psi} \sqrt{z} }{4 \epsilon^2} } \right) \right]
\end{aligned}
$$
Taking the limit yields:
$$
\begin{aligned}
	\mathcal{I}_{2a} &= \pi \left[ \delta\left( \hat{\psi} - i \frac{t}{2} \right) - \delta\left( \hat{\psi} + i \frac{t}{2} \right) + \frac{st}{2} \int_0^1 dz\, \frac{1}{ \sqrt{1-z} } I_1(st \sqrt{1-z}) \left( \delta\left( \hat{\psi} - i \frac{t \sqrt{z} }{2} \right) - \delta\left( \hat{\psi} + i \frac{t \sqrt{z} }{2} \right) \right) \right]
\end{aligned}
$$
Now for the second contribution:
$$
\begin{aligned}
	\mathcal{I}_{2b} &= \lim_{\epsilon \to 0} \frac{\sqrt{\pi}}{\epsilon} e^{ -\frac{1}{4} \frac{ \hat{\psi}^2 }{ \epsilon^2 } } \sum_{l=0}^\infty (-1)^l \left[ st \left( \frac{t}{2 \epsilon \sqrt{2}} \right)^{2l} H_{e_{2l}} \left( \frac{ \hat{\psi} }{ \sqrt{2} \epsilon } \right) \right] \frac{ \Gamma(l+3/2) }{ \Gamma(2l+2) } \sum_{k=0}^\infty \frac{1}{k! \Gamma(k+l+3/2)} \left( \frac{st}{2} \right)^{2k}
\end{aligned}
$$
Using the same identity and change of variables:
$$
\begin{aligned}
	\mathcal{I}_{2b} &= st \lim_{\epsilon \to 0} \frac{ \sqrt{\pi} }{2 \epsilon } e^{ -\frac{1}{4} \frac{ \hat{\psi}^2 }{ \epsilon^2 } } \int_0^1 \frac{dz}{ \sqrt{z} } \sum_{l=0}^\infty (-1)^l \left( \frac{t \sqrt{z} }{ 2 \epsilon \sqrt{2} } \right)^{2l} \frac{1}{\Gamma(2l+1)} 2^{-l} H_{2l} \left( \frac{ \hat{\psi} }{2\epsilon} \right) I_0( st \sqrt{1-z} )
\end{aligned}
$$
This becomes:
$$
\begin{aligned}
	\mathcal{I}_{2b} &= \frac{st}{2} \lim_{\epsilon \to 0} \frac{ \sqrt{\pi} }{ \epsilon } e^{ -\frac{1}{4} \frac{ \hat{\psi}^2 }{ \epsilon^2 } } \int_0^1 \frac{dz}{ \sqrt{z} } e^{ \frac{t^2 z}{16 \epsilon^2} } \cos\left( \frac{ t \hat{\psi} \sqrt{z} }{4 \epsilon^2} \right) I_0( st \sqrt{1-z} )
\end{aligned}
$$
Then, taking the limit:
$$
\begin{aligned}
	\mathcal{I}_{2b} &= \pi \frac{st}{2} \int_0^1 \frac{dz}{\sqrt{z}} I_0( st \sqrt{1-z} ) \left[ \delta\left( \hat{\psi} - i \frac{t \sqrt{z}}{2} \right) + \delta\left( \hat{\psi} + i \frac{t \sqrt{z}}{2} \right) \right]
\end{aligned}
$$
Combining both contributions, we obtain the final expression for $\mathcal{I}_2$:
\[
\mathcal{I}_2 = \pi \left[
\delta\left( \hat{\psi} - i \frac{t}{2} \right) - \delta\left( \hat{\psi} + i \frac{t}{2} \right)
+ \frac{st}{2} \int_0^1 dz \, \frac{1}{\sqrt{1 - z}} I_1(st \sqrt{1 - z}) \left( \delta\left( \hat{\psi} - i \frac{t \sqrt{z}}{2} \right) - \delta\left( \hat{\psi} + i \frac{t \sqrt{z}}{2} \right) \right) \right.
\]
\[
\left. + \frac{st}{2} \int_0^1 \frac{dz}{\sqrt{z}} I_0(st \sqrt{1 - z}) \left( \delta\left( \hat{\psi} - i \frac{t \sqrt{z}}{2} \right) + \delta\left( \hat{\psi} + i \frac{t \sqrt{z}}{2} \right) \right)
\right]
\]

\subsubsection{Going ahead to compute $P_{n,m}(t)$}
If we go back to equation (\ref{eq:pre_final_full}) and put everything together, $\mathcal{I}_1 + \mathcal{I}_{2a} + \mathcal{I}_{2b}$, we obtain:
$$
\begin{aligned}
	P_{n,m}(t) &= \frac{1}{2\pi} e^{ - (q + p + s)t } \int d\hat{\psi} \Bigg[
	2\pi\, \delta\left( \hat{\psi} - i \frac{t}{2} \right)
	+ \pi \frac{st}{2} \int_0^1 dz\, \frac{I_1\left( st \sqrt{1 - z} \right)}{ \sqrt{z(1 - z)} } \left( \delta\left( \hat{\psi} - i \frac{t}{2} \sqrt{z} \right) + \delta\left( \hat{\psi} + i \frac{t}{2} \sqrt{z} \right) \right) \\
	&\quad + \pi \frac{st}{2} \int_0^1 dz\, \frac{I_1\left( st \sqrt{1 - z} \right)}{ \sqrt{1 - z} } \left( \delta\left( \hat{\psi} - i \frac{t}{2} \sqrt{z} \right) - \delta\left( \hat{\psi} + i \frac{t}{2} \sqrt{z} \right) \right) \Bigg] \\
	&\quad \times I_{n - n_0} \left( \frac{q + p}{2}t - i\hat{\psi}(q - p) \right) I_{m - m_0} \left( \frac{q + p}{2}t + i\hat{\psi}(q - p) \right) \\
	&\quad + \frac{1}{2\pi} e^{ - (q + p + s)t } \int d\hat{\psi} \left[
	\pi \frac{st}{2} \int_0^1 dz\, \frac{I_0\left( st \sqrt{1 - z} \right)}{ \sqrt{z} }
	\left( \delta\left( \hat{\psi} - i \frac{t}{2} \sqrt{z} \right) + \delta\left( \hat{\psi} + i \frac{t}{2} \sqrt{z} \right) \right)
	\right] \\
	&\quad \times I_{n - m_0} \left( \frac{q + p}{2}t - i\hat{\psi}(q - p) \right)
	I_{m - n_0} \left( \frac{q + p}{2}t + i\hat{\psi}(q - p) \right)
\end{aligned}
$$
Now integrating over $\hat{\psi}$ using the delta functions, we get:
\begin{eqnarray}\label{ec.final_one}
	P_{n,m}(t) &=& e^{ - (q + p + s)t } \Bigg[\nonumber
	I_{n - n_0}(qt)\, I_{m - m_0}(pt)
	+ \sum_{\sigma = \pm 1} \frac{st}{4} \int_0^1 \frac{dz}{ \sqrt{z} }
	\left( \frac{1 + \sigma \sqrt{z}}{ \sqrt{1 - z} } I_1\left( st \sqrt{1 - z} \right) \right) \\
	&\times&  I_{n - n_0} \left( \frac{q + p}{2}t + \sigma \frac{t\sqrt{z}}{2}(q - p) \right)
	I_{m - m_0} \left( \frac{q + p}{2}t - \sigma \frac{t\sqrt{z}}{2}(q - p) \right)
	\Bigg] \nonumber\\
	&+&  e^{ - (q + p + s)t } \sum_{\sigma = \pm 1} \frac{st}{4} \int_0^1 \frac{dz}{ \sqrt{z} }
	I_0\left( st \sqrt{1 - z} \right)
	I_{n - m_0} \left( \frac{q + p}{2}t + \sigma \frac{t\sqrt{z}}{2}(q - p) \right)
	I_{m - n_0} \left( \frac{q + p}{2}t - \sigma \frac{t\sqrt{z}}{2}(q - p) \right)\nonumber
	\\&&
\end{eqnarray}

This expression gives the full solution in equation (\ref{eq:fullsolution}).

A simple consistency check—namely that the total probability is conserved, i.e.,
$
\sum_{n,m} P_{n,m}(t) = 1,
$
can be done using identity {\bf 6.673-3} from Gradhsteyn and Ryzhik.

\subsection{Marginal Distributions}

The marginal probability densities in each channel are obtained by summing over the spatial coordinate in the other channel:
\[
P^1_{n}(t) = \sum_m P_{n,m}(t), \qquad P^2_{m}(t) = \sum_n P_{n,m}(t).
\]
These represent the probability of finding the particle at site \(n\) in channel 1, or at site \(m\) in channel 2, at time \(t\), regardless of its position in the other channel. From the exact expression for \(P_{n,m}(t)\), we obtain:
\begin{align}
	P^1_{n}(t) &= e^{-(q+s)t}\,I_{n-n_0}(qt) + e^{-\left(\tfrac{q+p}{2} + s\right)t} \sum_{\sigma=\pm1} \frac{st}{4} \int_0^1 \frac{dz}{\sqrt{z}}\, \frac{1 + \sigma\sqrt{z}}{\sqrt{1 - z}}\, I_1\bigl(st\sqrt{1 - z}\bigr) \nonumber \\
	&\quad \times I_{n - n_0}\left(\tfrac{q+p}{2}t + \sigma \tfrac{t\sqrt{z}}{2}(q - p)\right)\, e^{-\sigma \tfrac{t\sqrt{z}}{2}(q - p)} \nonumber \\
	&\quad + e^{-\left(\tfrac{q+p}{2} + s\right)t} \sum_{\sigma=\pm1} \frac{st}{4} \int_0^1 \frac{dz}{\sqrt{z}}\, I_0\bigl(st\sqrt{1 - z}\bigr)\, I_{n - m_0}\left(\tfrac{q+p}{2}t + \sigma \tfrac{t\sqrt{z}}{2}(q - p)\right)\, e^{-\sigma \tfrac{t\sqrt{z}}{2}(q - p)} \label{eq:Pn_marg},
\end{align}
\begin{align}
	P^2_{m}(t) &= e^{-(p+s)t}\,I_{m - m_0}(pt) + e^{-\left(\tfrac{q+p}{2} + s\right)t} \sum_{\sigma=\pm1} \frac{st}{4} \int_0^1 \frac{dz}{\sqrt{z}}\, \frac{1 - \sigma\sqrt{z}}{\sqrt{1 - z}}\, I_1\bigl(st\sqrt{1 - z}\bigr) \nonumber \\
	&\quad \times I_{m - m_0}\left(\tfrac{q+p}{2}t + \sigma \tfrac{t\sqrt{z}}{2}(q - p)\right)\, e^{-\sigma \tfrac{t\sqrt{z}}{2}(q - p)} \nonumber \\
	&\quad + e^{-\left(\tfrac{q+p}{2} + s\right)t} \sum_{\sigma=\pm1} \frac{st}{4} \int_0^1 \frac{dz}{\sqrt{z}}\, I_0\bigl(st\sqrt{1 - z}\bigr)\, I_{m - n_0}\left(\tfrac{q+p}{2}t + \sigma \tfrac{t\sqrt{z}}{2}(q - p)\right)\, e^{-\sigma \tfrac{t\sqrt{z}}{2}(q - p)} \label{eq:Pm_marg}.
\end{align}

Notably, the mathematical structure of the marginal distributions reveals an important connection to the switching diffusion processes discussed in the introduction. If we consider the sum of both marginal distributions, $P^1_n(t) + P^2_n(t)$, and focus specifically on the terms containing $I_{n-n_0}(\cdot)$ (those preserving the initial identity starting at $n_0$), we recover precisely the probability distribution for a single particle that stochastically switches between two diffusive states with rates $p$ and $q$ at switching rate $s$. The exact equation for this switching random walk is then given by
\begin{align}
	p_{n}(t) &= e^{-(q+s)t}\,I_{n-n_0}(qt) + e^{-\left(\tfrac{q+p}{2} + s\right)t} \sum_{\sigma=\pm1} \frac{st}{4} \int_0^1 \frac{dz}{\sqrt{z}}\, \frac{1 + \sigma\sqrt{z}}{\sqrt{1 - z}}\, I_1\bigl(st\sqrt{1 - z}\bigr) \nonumber \\
	&\quad \times I_{n - n_0}\left(\tfrac{q+p}{2}t + \sigma \tfrac{t\sqrt{z}}{2}(q - p)\right)\, e^{-\sigma \tfrac{t\sqrt{z}}{2}(q - p)} \nonumber \\
	&\quad + e^{-\left(\tfrac{q+p}{2} + s\right)t} \sum_{\sigma=\pm1} \frac{st}{4} \int_0^1 \frac{dz}{\sqrt{z}}\, I_0\bigl(st\sqrt{1 - z}\bigr)\, I_{n - n_0}\left(\tfrac{q+p}{2}t + \sigma \tfrac{t\sqrt{z}}{2}(q - p)\right)\, e^{-\sigma \tfrac{t\sqrt{z}}{2}(q - p)}.
\end{align}
This formulation connects directly to the intermittent active motion models discussed in our introduction \cite{Datta24}. Crucially, the terms with $I_{n-n_0}$ in $P_n^1(t)$ correspond to the particle maintaining its original identity, while those in $P_n^2(t)$ represent the swapped configuration, exactly analogous to state transitions in switching diffusion. This full discrete-random-walk formulation for two walkers is, to our knowledge, absent from existing discrete-walk literature. A comparable continuous analogue was developed by Lee \& Hill (1982) in their study \textit{On the General Linear Coupled System for Diffusion in Media with Two Diffusivities}.

However, the full joint distribution $P_{n,m}(t)$ cannot be reconstructed from the marginals as a product $P^1_{n}(t)P^2_{m}(t)$, as the marginal densities alone would only describe two particles switching states independently at rate $s$. In the actual joint process, the two components evolve together under shared switching dynamics, which induces correlations between $n$ and $m$ and captures the synchronous exchange that is lost when considering the marginals separately. These correlations, absent in a product of marginals, encode the nontrivial structure of inter-channel memory and mutual dependence that the model captures exactly, distinguishing it from independent switching diffusion processes.

  \subsection{Limiting cases}
 \subsubsection{Short time behavior according to $s$, i.e., $st \rightarrow 0$}
 This case is straightforward. When $st \rightarrow 0$, the integral contributions vanish, and the result reduces to the simple, uncoupled dynamics:
 $$
 P_{n,m}(t) = e^{-(q + p)t} \, I_{n - n_0}(qt) \, I_{m - m_0}(pt),
 $$
 as expected.
 
 \subsubsection{Long time behavior according to $s$, i.e., $st \rightarrow \infty$} 
 This case is more subtle. 
 In the large-\(t\) regime with \(st \gg 1\), the first term (a direct product of Bessel functions) is exponentially suppressed as \(e^{-st}\) and can be neglected. The two integral terms, denoted \(T_1\) and \(T_2\), dominate in this limit.
 
 We approximate the Bessel functions \(I_0\) and \(I_1\) using their leading asymptotic forms for large argument:
 \[
 \small
 I_k(z) \sim \frac{e^z}{\sqrt{2\pi z}}, \quad z \to \infty.
 \]
 In the integrands, the arguments of \(I_0\) and \(I_1\) contain the factor \(st \sqrt{1 - z}\), which is large for all \(z \in [0, 1)\) as \(st \to \infty\). Thus,
 \[
 \small
 I_{0,1}\left( st \sqrt{1 - z} \right) \sim \frac{e^{st \sqrt{1 - z}}}{\sqrt{2\pi st}(1 - z)^{1/4}}.
 \]
 Since the exponential term \(e^{st \sqrt{1 - z}}\) peaks at \(z = 0\), we perform a local expansion around \(z = 0\). Approximating \(\sqrt{1 - z} \approx 1 - z/2\) and \((1 - z)^{-1/4} \approx 1\), we write:
 \[
 \small
 e^{st \sqrt{1 - z}} \approx e^{st} e^{-st z / 2}.
 \]
 The remaining algebraic prefactors are regular at \(z = 0\), and the integral becomes dominated by small-\(z\) contributions. We also approximate the Bessel functions involving \(n\), \(m\), \(n_0\), and \(m_0\) at \(z = 0\), yielding:
 \[
 \small
 \begin{aligned}
 	I_{n - n_0} &\left( \tfrac{q + p}{2}t + \sigma \tfrac{t\sqrt{z}}{2}(q - p) \right) \approx I_{n - n_0}\left( \tfrac{q + p}{2}t \right), \\
 	I_{m - m_0} &\left( \tfrac{q + p}{2}t - \sigma \tfrac{t\sqrt{z}}{2}(q - p) \right) \approx I_{m - m_0}\left( \tfrac{q + p}{2}t \right),
 \end{aligned}
 \]
 and similarly for the terms in \(T_2\). Thus, we obtain the simplified approximations:
 \[
 \small
 \begin{aligned}
 	T_1 &\sim e^{-(q + p + s)t} \sum_{\sigma = \pm 1} \frac{st}{4} \int_0^1 \frac{dz}{\sqrt{z}} \frac{1 + \sigma \sqrt{z}}{(1 - z)^{3/4}} \frac{e^{st \sqrt{1 - z}}}{\sqrt{2\pi st}} F(0), \\
 	T_2 &\sim e^{-(q + p + s)t} \sum_{\sigma = \pm 1} \frac{st}{4} \int_0^1 \frac{dz}{\sqrt{z}} \frac{e^{st \sqrt{1 - z}}}{\sqrt{2\pi st} (1 - z)^{1/4}} G(0),
 \end{aligned}
 \]
 where
 \[
 \small
 \begin{aligned}
 	F(0) &= I_{n - n_0}\left( \tfrac{q + p}{2}t \right) I_{m - m_0}\left( \tfrac{q + p}{2}t \right), \\
 	G(0) &= I_{n - m_0}\left( \tfrac{q + p}{2}t \right) I_{m - n_0}\left( \tfrac{q + p}{2}t \right).
 \end{aligned}
 \]
 Using the exponential localization of the integrand around \(z = 0\), we extend the integrals to \([0, \infty)\) and evaluate:
 \[
 \small
 \int_0^\infty \frac{dz}{\sqrt{z}} e^{-st z / 2} = \sqrt{\frac{2\pi}{st}}, \quad 
 \int_0^\infty \frac{dz}{\sqrt{z}} \sqrt{z} e^{-st z / 2} = \frac{2}{st}.
 \]
 Combining terms and summing over \(\sigma\), we find:
 \[
 \small
 T_1 \sim e^{-(q + p)t} \cdot \frac{1}{2} F(0), \quad
 T_2 \sim e^{-(q + p)t} \cdot \frac{1}{2} G(0).
 \]
 Thus, the total asymptotic behavior of the probability distribution is
 \begin{equation}\label{ec:long_time_density}
  	P_{n,m}(t) \approx \frac{e^{-(q+p)t}}{2} \left[ I_{n - n_0}\left( \tfrac{q + p}{2}t \right) I_{m - m_0}\left( \tfrac{q + p}{2}t \right) + I_{n - m_0}\left( \tfrac{q + p}{2}t \right) I_{m - n_0}\left( \tfrac{q + p}{2}t \right) \right]
\end{equation}
 This result holds for fixed \(n, m, n_0, m_0\) in the limit \(t \to \infty\) with \(st \to \infty\). The final expression is symmetric under exchange of the initial coordinates \(n_0 \leftrightarrow m_0\), reflecting the underlying reversibility and coupling structure of the stochastic dynamics. In the special case \(n_0 = m_0 = 0\), this reduces to:
 \[
 P_{n,m}(t) \approx e^{-(q + p)t} I_n\left( \tfrac{q + p}{2}t \right) I_m\left( \tfrac{q + p}{2}t \right).
 \]

 \subsubsection*{Long time behavior according to $p+q$ and $s$, i.e., (p+q)t$\rightarrow\infty$  and st$\rightarrow\infty$}
 
 Having identified the dominant contribution to \(P_{n,m}(t)\) in the regime \(st \to \infty\) in equation~\eqref{ec:long_time_density}, we now proceed to analyze its asymptotic behavior in the limit \(t \to \infty\), keeping \(n\), \(m\), \(n_0\), and \(m_0\) fixed. In this regime, the Bessel functions \(I_k\left( \frac{q + p}{2} t \right)\) appearing in \eqref{ec:long_time_density} can be approximated by their uniform second-order expansion~\cite{olver1974asymptotics,temme1996special}, valid when \(k = \mathcal{O}(t^{1/2})\). Explicitly,
 \[
 I_k(\alpha) = \frac{e^{\alpha}}{\sqrt{2\pi\alpha}} \exp\left( -\frac{k^2}{2\alpha} \right) \left[ 1 + \frac{1}{8\alpha} \left( 1 - \frac{k^2}{\alpha} \right) + \mathcal{O}(\alpha^{-2}) \right],
 \]
 where \(\alpha = \frac{(p+q)t}{2}\). This expansion is obtained by applying Laplace’s method to the integral representation of 
 $I_k(\alpha) = \frac{1}{2\pi} \int_{-\pi}^{\pi} e^{\alpha \cos\theta - ik\theta} d\theta
 $ and remains valid for fixed \(k\) or slowly growing \(k \ll \alpha^{3/4}\). The derivation proceeds by expanding the exponent \(\alpha \cos \theta - i k \theta\) around its maximum at \(\theta = 0\), leading to a Gaussian integral once the integrand is approximated as \(e^{\alpha - \frac{\alpha}{2} \theta^2 - i k \theta}\). Evaluating this integral yields the leading-order term \(\frac{e^\alpha}{\sqrt{2\pi\alpha}} \exp(-k^2/2\alpha)\), while the correction terms follow from a higher-order expansion of the exponent. This asymptotic form captures the sharp peak and decay of \(I_k(\alpha)\) for large arguments and is rigorously derived in standard references such as~\cite{olver1974asymptotics,temme1996special}.
 
 We now use this expansion to approximate each term in \eqref{ec:long_time_density}. Introducing the center of mass \(c = \frac{n_0 + m_0}{2}\) and the initial asymmetry \(\delta = n_0 - m_0\), the expression becomes:
 \[
 \begin{aligned}
 	P_{n,m}(t) &\approx \frac{e^{-(q + p)t} e^{(q + p)t}}{4\pi\alpha} \Bigg\{
 	\exp\left[ -\frac{(n - n_0)^2 + (m - m_0)^2}{2\alpha} \right]
 	+
 	\exp\left[ -\frac{(n - m_0)^2 + (m - n_0)^2}{2\alpha} \right]
 	\Bigg\} \\
 	&\quad \times \left[ 1 + \mathcal{O}(t^{-1}) \right].
 \end{aligned}
 \]
 
 Expanding the exponents and reorganizing in terms of \(c\) and \(\delta\), we find
 \[
 \frac{(n - n_0)^2 + (m - m_0)^2}{2\alpha} = \frac{(n - c)^2 + (m - c)^2}{(p + q)t} + \frac{\delta^2}{4(p + q)t} + \frac{(n - m)\delta}{2(p + q)t},
 \]
 and analogously for the second term with \(\delta \to -\delta\). Summing both contributions and using the identity \(\cosh(z) = \frac{1}{2}(e^z + e^{-z})\), we obtain:
 \[
 P_{n,m}(t) \approx \frac{1}{\pi(p + q)t} \exp\left( -\frac{(n - c)^2 + (m - c)^2}{(p + q)t} - \frac{\delta^2}{4(p + q)t} \right) \cosh\left( \frac{(n - m)\delta}{2(p + q)t} \right) + \mathcal{O}(t^{-2}).
 \]
 
 This form reveals that for large \(t\), the probability distribution becomes sharply peaked around the center of mass \(c\), modulated by a weak dependence on the initial asymmetry \(\delta\) and the difference \(n - m\). The correction encoded in the \(\cosh\) term becomes negligible when its argument is small. Assuming diffusive scaling \(n - m \sim \sqrt{(p + q)t}\), the argument of the hyperbolic cosine becomes small when
 \[
 \left| \frac{(n - m)\delta}{2(p + q)t} \right| \ll 1 \quad \Longrightarrow \quad t \gg \frac{\delta^2}{4(p + q)^2}.
 \]
 This suggests the existence of a characteristic timescale
 \[
 t_{\text{crit}} = \frac{(n_0 - m_0)^2}{4(p + q)^2},
 \]
 beyond which the memory of the initial asymmetry is lost and the distribution becomes approximately symmetric.
 
 Finally, for \(t \gg t_{\text{crit}}\), we can set \(\cosh(\cdot) \approx 1\) and neglect the exponential prefactor involving \(\delta^2\), leading to the asymptotic expression
 \[
 P_{n,m}(t) \approx \frac{1}{\pi(p + q)t} \exp\left( -\frac{(n - c)^2 + (m - c)^2}{(p + q)t} \right) + \mathcal{O}(t^{-3/2}).
 \]
 This is a symmetric product of Gaussians centered at the average initial position, emerging from the convolution of two Bessel functions at large time.

 \subsubsection*{Marginal density in the long-time regime: $(p+q)t \to \infty$ and $st \to \infty$}
 
 We now turn to the asymptotic behavior of the marginal distribution \( P^1_{n}(t) = \sum_m P_{n,m}(t) \) in the regime where both \( (p+q)t \to \infty \) and \( st \to \infty \), using the asymptotic structure already established for the joint distribution in equation~\eqref{ec:long_time_density}. To proceed, we isolate the dependence on the index \( n \) by analyzing the summation over Bessel functions. Specifically, we focus on approximating terms of the form \( I_{n - n_0}(x) + I_{n - m_0}(x) \), with \( x = \frac{(p+q)t}{2} \), in the limit \( x \gg 1 \).
 
 Letting \(c = \frac{n_0 + m_0}{2}\), \(\delta = \frac{n_0 - m_0}{2}\), and \(\nu = n - c\), we rewrite the sum as
 \[
 I_{n - n_0}(x) + I_{n - m_0}(x) = I_{\nu - \delta}(x) + I_{\nu + \delta}(x).
 \]
 Expanding around \(\delta = 0\), a Taylor expansion yields:
 \[
 I_{\nu \pm \delta}(x) = I_\nu(x) \pm \delta \frac{\partial I_\nu}{\partial \nu} + \frac{\delta^2}{2} \frac{\partial^2 I_\nu}{\partial \nu^2} + \mathcal{O}(\delta^3).
 \]
 Adding both terms and retaining terms up to \(\mathcal{O}(\delta^2)\), we obtain
 \[
 I_{\nu - \delta}(x) + I_{\nu + \delta}(x) = 2 I_\nu(x) + \delta^2 \frac{\partial^2 I_\nu}{\partial \nu^2} + \mathcal{O}(\delta^4).
 \]
 
 Using the known asymptotic expansion for \(I_\nu(x)\) in the regime \(x \gg |\nu|\) and \(x \gg 1\), we have:
 \[
 I_\nu(x) \sim \frac{e^x}{\sqrt{2\pi x}}\left(1 - \frac{4\nu^2 - 1}{8x} + \mathcal{O}(x^{-2})\right),
 \]
 from which it follows that
 \[
 \frac{\partial^2 I_\nu}{\partial \nu^2} \sim I_\nu(x)\left(-\frac{1}{x} + \frac{\nu^2}{x^2} + \mathcal{O}(x^{-2})\right).
 \]
 Therefore, combining the terms, we find
 \[
 I_{n - n_0}(x) + I_{n - m_0}(x) = 2 I_{n - c}(x)\left(1 - \frac{(n_0 - m_0)^2}{8x}\right) + \mathcal{O}\left(\frac{(n_0 - m_0)^4}{x^2}, \frac{1}{x^2}\right).
 \]
 
 Substituting this result into equation~\eqref{ec:long_time_density}, and summing over \(m\) to obtain the marginal distribution, we get
 \[
 P^1_{n}(t) = \sum_m P_{n,m}(t) \approx \frac{e^{-(p+q)t}}{2} \left[ I_{n - n_0}(x) + I_{n - m_0}(x) \right] \sum_m I_{m - m_0}(x),
 \]
 where the sum over \(m\) of the Bessel function yields
 \[
 \sum_m I_{m - m_0}(x) = e^x,
 \]
 due to the identity \( \sum_k I_k(x) = e^x \). Simplifying, we obtain
 \[
 P^1_{n}(t) \approx \frac{1}{2}  e^{-\frac{(p+q)t}{2}} \cdot \left[ I_{n - n_0}(x) + I_{n - m_0}(x) \right].
 \]
 
 Using the expansion derived above, the marginal probability becomes:
 \[
 P^1_{n}(t) \approx e^{-(p+q)t/2} \cdot I_{n - c}(x)\left(1 - \frac{(n_0 - m_0)^2}{8x}\right) + \mathcal{O}(t^{-2}),
 \]
 where \(x = \frac{(p+q)t}{2}\). Thus, the marginal distribution takes the form:
 \[
 	P^1_{n}(t) \approx e^{-(p+q)t/2} \cdot I_{n - c}\left( \tfrac{(p+q)t}{2} \right) \left(1 - \frac{(n_0 - m_0)^2}{4(p+q)t} \right) + \mathcal{O}(t^{-2})
 \]
 
 This result highlights that the marginal density inherits the Gaussian-like spreading of the Bessel function centered at the average initial position \(c\), with a small correction due to the initial asymmetry \((n_0 - m_0)\). The correction decays as \(1/t\), indicating that at long times, the marginal distribution loses memory of the initial imbalance and converges to a single-locus diffusion profile.

%

	\subsection{Moments}
	
	The statistical moments of the joint process can be computed either directly from the probability distribution \( P_{n,m}(t) \) or more efficiently via the Fourier-transformed propagator
	\begin{equation}
		G(k_x,k_y,t) = \sum_{n,m} e^{i k_x n + i k_y m}\,P_{n,m}(t),
	\end{equation}
	which corresponds to the two-dimensional characteristic function. In this representation, the moments are obtained by differentiating with respect to the Fourier variables:
	\begin{eqnarray}
		\left.\frac{\partial G}{\partial k_x}\right|_{k=0} &=& i\,\langle n(t) \rangle, \qquad\quad
		\left.\frac{\partial G}{\partial k_y}\right|_{k=0} = i\,\langle m(t) \rangle, \\
		\left.\frac{\partial^2 G}{\partial k_x^2}\right|_{k=0} &=& -\langle n(t)^2 \rangle, \qquad\
		\left.\frac{\partial^2 G}{\partial k_y^2}\right|_{k=0} = -\langle m(t)^2 \rangle, \\
		\left.\frac{\partial^2 G}{\partial k_x\,\partial k_y}\right|_{k=0} &=& -\langle n(t)\,m(t) \rangle.
	\end{eqnarray}
	Using the exact expression for \( G(k_x, k_y, t) \) given in Eq.~(\ref{eq:sol2D-Laplace}), the following results are obtained for the first and second moments of the walker positions:
	\begin{eqnarray}
		\langle n(t) \rangle &=& \frac{n_0 + m_0}{2} + \frac{(n_0 - m_0)}{2}\,e^{-2s t}, \\
		\langle m(t) \rangle &=& \frac{n_0 + m_0}{2} + \frac{(m_0 - n_0)}{2}\,e^{-2s t},
	\end{eqnarray}
	\begin{eqnarray}
		\langle n(t)^2 \rangle &=& \frac{(1 - e^{-2s t})(q - p)}{4s} + \frac{(q + p)t}{2} + \frac{(m_0^2 + n_0^2)}{2} + \frac{(n_0^2 - m_0^2)}{2}\,e^{-2s t}, \\
		\langle m(t)^2 \rangle &=& \frac{(1 - e^{-2s t})(p - q)}{4s} + \frac{(q + p)t}{2} + \frac{(m_0^2 + n_0^2)}{2} + \frac{(m_0^2 - n_0^2)}{2}\,e^{-2s t},
	\end{eqnarray}
	\begin{equation}
		\langle n(t)\,m(t) \rangle = n_0\,m_0.
	\end{equation}
	From these, the variances are:
	\begin{eqnarray}
		 \langle n(t)^2 \rangle_v &=& \langle n(t)^2 \rangle - \langle n(t) \rangle^2 = \frac{(n_0 - m_0)^2}{4}\left(1 - e^{-4s t}\right) + \frac{(q - p)(1 - e^{-2s t})}{4s} + \frac{(q + p)t}{2}, \\
		\langle m(t)^2 \rangle_v &=& \langle m(t)^2 \rangle - \langle m(t) \rangle^2 = \frac{(n_0 - m_0)^2}{4}\left(1 - e^{-4s t}\right) + \frac{(p - q)(1 - e^{-2s t})}{4s} + \frac{(q + p)t}{2},
	\end{eqnarray}
	with the difference between variances given by:
	\begin{equation}
		\langle n(t)^2 \rangle_v - \langle m(t)^2 \rangle_v = \frac{(q - p)(1 - e^{-2s t})}{2s}.
	\end{equation}
	Finally, the covariance between the two channels is:
	\begin{equation}
		\langle n(t)\,m(t) \rangle_c = \langle n(t)\,m(t) \rangle - \langle n(t) \rangle \langle m(t) \rangle = -\frac{(n_0 - m_0)^2}{4}\left(1 - e^{-4s t}\right).
	\end{equation}
	Note that the covariance vanishes at \( t = 0 \) and decays monotonically as \( t \to \infty \), indicating the emergence of correlations between the two components solely due to the inter-channel coupling. All moment expressions presented here follow directly from exact computation using the generating function approach, though they may alternatively be obtained by integration over \( P_{n,m}(t) \) with appropriate weights.
	
	\subsection{Time Correlation Function}
	To compute the time correlation function of the position process \( n(t) \), we consider the two-time expectation
	\begin{equation}
		B_n(t, t + \tau) = \langle n(t)\,n(t+\tau) \rangle = \sum_{n,n'} n\,n'\,p(n,t;\,n',t+\tau),
	\end{equation}
	where \( p(n,t;\,n',t+\tau) \) denotes the joint probability of the walker being at site \( n \) at time \( t \), and at site \( n' \) at time \( t + \tau \).
	By conditioning on the intermediate value and using the marginalization \( p(n,t) = \sum_m P_{n,m}(t) \), we write:
	\begin{align}
		B_n(t, t+\tau) &= \sum_{n,n'} n\,n'\,p(n', t+\tau \mid n, t)\,p(n, t) \notag \\
		&= \sum_{n,m} n\,p(n,m,t \mid n_0, m_0) \sum_{n',m'} n'\,p(n', m', t+\tau \mid n, m, t). 
	\end{align}
	From the exact expression for the conditional expectation $E[n(t+\tau)|n(t)=n,m(t)=m]$, we have:
	\begin{equation}
		\sum_{n',m'} n'\,p(n', m', t+\tau \mid n, m, t) = \frac{n + m}{2} + \frac{(n - m)}{2}e^{-2s\tau}.
	\end{equation}
	Substituting this into the sum, we obtain:
	\begin{align}
		B_n(t, t+\tau) &= \sum_{n,m} n\,p(n,m,t) \left[ \frac{n + m}{2} + \frac{(n - m)}{2}e^{-2s\tau} \right] \notag \\
		&= \frac{1 + e^{-2s\tau}}{2} \langle n(t)^2 \rangle + \frac{1 - e^{-2s\tau}}{2} \langle n(t)\,m(t) \rangle.
	\end{align}
	Using the explicit expressions for \( \langle n(t)^2 \rangle \) and \( \langle n(t)\,m(t) \rangle \) from the previous section, this yields:
	\begin{align}
		B_n(t, t+\tau) &= \frac{1 + e^{-2s\tau}}{2} \left[ \frac{(1 - e^{-2s t})(q - p)}{4s} + \frac{(p + q)t}{2} + \frac{e^{-2s t}(n_0^2 - m_0^2)}{2} + \frac{m_0^2 + n_0^2}{2} \right] \notag \\
		&\quad + \frac{1 - e^{-2s\tau}}{2} \cdot m_0 n_0.
	\end{align}
	
	\subsection{Diffusion Exponent}
	
	We define the diffusion exponent \( \alpha_{n,m}(t) \) via the scaling relation:
	\begin{equation}
		\alpha_{n}(t) = \frac{d\log  \langle n(t)^2 \rangle_v}{d\log t}, \qquad 	\alpha_{m}(t) = \frac{d\log  \langle m(t)^2 \rangle_v}{d\log t},
	\end{equation}
	which can be computed explicitly using the exact variance expressions. Introducing the auxiliary parameters:
	\[
	\theta = \frac{(m_0 - n_0)^2}{q + p}, \qquad \rho = \frac{q - p}{q + p},
	\]
	we obtain the time-dependent diffusion exponent in closed form:
	\begin{equation}
		\alpha_{n,m}(t) =
		\frac{1 + 2s\theta\,e^{-4st} \pm \rho\,e^{-2st}}{
			1 - \frac{s\theta}{2st}(1 - e^{-4st}) \pm \frac{\rho}{2st}(1 - e^{-2st}) },
	\end{equation}
	where the upper sign corresponds to \( \alpha_n(t) \) and the lower to \( \alpha_m(t) \).
	This function quantifies the crossover from anomalous to normal diffusion at long times, with the asymptotic limit \( \alpha_{n,m}(t) \to 1 \) as \( t \to \infty \).
	
	In the particular case where the initial conditions satisfy $n_0 = m_0$, the diffusion exponent simplifies to
	\begin{equation}
		\alpha_{n,m}(t) = \frac{1 \pm \frac{q-p}{q+p} e^{-2st}}{1 \pm \frac{q-p}{q+p} \frac{1 - e^{-2st}}{2 s t}},
	\end{equation}
	recovering the expression previously given in equation (\ref{eq.diff_coef_n00}). This form highlights the pure effect of the asymmetry ratio $\frac{q-p}{q+p}$ on the diffusion scaling without the influence of the initial position difference.

\end{document}